\documentclass[prl,aps,twocolumn,%
showpacs,preprintnumbers,amsmath,amssymb,nofootinbib]{revtex4-2}
\usepackage[utf8]{inputenc}

\usepackage{graphicx}
\usepackage{bm}
\usepackage[breaklinks, colorlinks=true, pdfstartview=FitV, citecolor=blue, urlcolor=blue]{hyperref}
\usepackage{slashed}
\usepackage{array}
\usepackage{mathtools}

\newcommand{\diff}{\mathrm{d}}

\newcommand{\tr}{\mathrm{tr}}
\newcommand{\im}{\mathrm{i}}

\newcommand{\rme}{\mathrm{e}}

\newcommand{\zetam}{\zeta_{\mathrm{m}}}
\newcommand{\zetav}{\zeta_{\mathrm{v}}}

\begin{document}

\preprint{YITP-24-59}

\title{Unifying Monopole and Center Vortex as the Semiclassical Confinement Mechanism}

\author{Yui Hayashi}
\email{yui.hayashi@yukawa.kyoto-u.ac.jp} 
\affiliation{Yukawa Institute for Theoretical Physics,  Kyoto University, Kyoto 606-8502, Japan}

\author{Yuya Tanizaki}
\email{yuya.tanizaki@yukawa.kyoto-u.ac.jp}
\affiliation{Yukawa Institute for Theoretical Physics,  Kyoto University, Kyoto 606-8502, Japan}

\begin{abstract}
Magnetic excitations play a crucial role in understanding the color confinement of $4$d Yang-Mills theory, and we have the monopole and the center vortex as plausible candidates to explain its mechanism. 
Under suitable compactified setups of $4$d Yang-Mills theory, we can achieve different weakly-coupled descriptions of confinement phenomena: The monopole mechanism takes place on $\mathbb{R}^3\times S^1$ with the double-trace deformation, and the center-vortex mechanism is effective on $\mathbb{R}^2\times T^2$ with the 't~Hooft flux. 
We unify these two semiclassical descriptions by showing the explicit relation between the monopole and center vortex. 
\end{abstract}

\maketitle
\section{Introduction}

Understanding color confinement still lies in front of us as a longstanding problem in particle and nuclear physics. 
More than 15 years ago, there was an important finding on this problem, which uncovers that qualitative aspects of confinement are kept intact in a suitable deformation of the $4$d Yang-Mills (YM) theory on $\mathbb{R}^3\times S^1$~\cite{Unsal:2007vu, Unsal:2007jx, Unsal:2008ch}. 
Starting from the $4$d YM Lagrangian with the center-stabilizing potential, we can derive the $3$d effective theory as the gas of fundamental monopoles in an analytical manner, and it allows us to access nonperturbative information of $4$d gauge theories using the weak-coupling description~\cite{Unsal:2007vu, Unsal:2007jx, Unsal:2008ch, Shifman:2008ja, Poppitz:2012nz, Poppitz:2012sw, Davies:2000nw}.  

Recently, there has been another development for the semiclassical description of the confinement phenomena by considering the YM theory on $\mathbb{R}^2\times T^2$ with the 't~Hooft flux~\cite{Tanizaki:2022ngt, Tanizaki:2022plm, Hayashi:2023wwi, Hayashi:2024qkm, Hayashi:2024gxv} (see~\cite{Yamazaki:2017ulc, Cox:2021vsa, Poppitz:2022rxv, Anber:2023sjn} for related studies). 
We can reduce the $4$d YM theory to the $2$d effective theory described by the gas of center vortices, and this unveils properties of confinement of the YM theory and chiral symmetry breaking of quantum chromodynamics (QCD). 

In both setups, the $4$d instanton splits into $N$ constituents carrying color magnetic fields: On $\mathbb{R}^3\times S^1$, the instanton consists of $N$ distinct fundamental monopoles, while on $\mathbb{R}^2\times T^2$ with the 't~Hooft flux, the instanton decomposes into $N$ identical center vortices. 
Therefore, we now have two distinct semiclassical theories of confinement for $4$d non-Abelian gauge theories. 
In the following of this Letter, we unify these two semiclassical theories of confinement after a brief review of each description. 

\section{Confinement on \texorpdfstring{$\mathbb{R}^3\times S^1$}{R3xS1} and monopoles}

Let us put the $4$d YM theory on $\mathbb{R}^3\times S^1 \ni (\vec{x},x_4)$ with $x_4\sim x_4 + L_4$, and we assume that $L_4$ is much smaller than the dynamical scale $\Lambda$, $NL_4 \Lambda\ll 1$. 
The one-loop effective potential~\cite{Gross:1980br} for the Wilson loop along the $4$th direction, $P_4(\vec{x})=\mathcal{P}\exp(\im\oint a_4(\vec{x},x_4)\diff x_4)$, prefers the center-broken vacua, $P_4=\rme^{2\pi \im m/N}\bm{1}_N$ ($m=1,\ldots, N$), so we need to add a center-stabilizing potential to realize the confinement phase in this setup~\cite{Unsal:2007jx, Unsal:2007vu, Unsal:2008ch}. 
To this end, we add the double-trace deformation to the effective potential, $\sum_{m=1}^{N-1} J_m |\tr(P_4^m)|^2$ with $J_m>0$. Up to the gauge transformation, the vacuum of this potential is given by the $SU(N)$ clock matrix,
\begin{equation}
    P_4\propto C = \mathrm{diag}(1,\omega, \ldots, \omega^{N-1}), 
    \label{eq:VEVofP4}
\end{equation}
with $\omega=\rme^{2\pi \im/N}$, and we work on this gauge fixing. As a result, the low-energy gauge group of the $3$d effective theory becomes $SU(N)\xrightarrow{\mathrm{Higgs}}U(1)^{N-1}$. 

The $3$d $U(1)$ gauge field is equivalent to the compact boson via Abelian duality. 
As a result, the $4$d YM theory on $\mathbb{R}^3\times S^1$ with the double-trace deformation can be described by the theory of dual photons $\vec{\sigma}(\vec{x})$, and it is related to the $U(1)^{N-1}$ field strength by 
\begin{equation}
    \vec{f}=\frac{g^2}{4\pi \im L_4}\star \diff \vec{\sigma}.
    \label{eq:DualPhotonDef}
\end{equation}
The periodicity of the dual photon is given by the fundamental weight vectors $\vec{\mu}_n$ for $\mathfrak{su}(N)$:\footnote{Our convention for the weight vectors is as follows: Take the orthonormal basis $\{\vec{e}_n\}_{n=1}^{N}$ of $\mathbb{R}^N$, and the simple roots are given by $\vec{\alpha}_n=\vec{e}_n-\vec{e}_{n+1}$ $(n=1, \ldots, N-1)$. The fundamental weights are $\vec{\mu}_n=\vec{e_1}+\cdots+\vec{e}_n-\frac{n}{N}\sum_{k=1}^{N}\vec{e}_k$, so that $\vec{\mu}_n\cdot \vec{\alpha}_m=\delta_{nm}$. }
\begin{equation}
    \vec{\sigma}(\vec{x})\sim \vec{\sigma}(\vec{x})+2\pi \vec{\mu}_n
    \quad (n=1,\ldots, N-1). 
    \label{eq:DualPhotonPeriodicity}
\end{equation}
We can show that the $4$d instanton splits into $N$ fundamental monopoles in this setup: 
$N-1$ of them are called the Bogomolny–Prasad–Sommerfield (BPS) monopoles, which carry the fractional topological charge $1/N$ and the magnetic charge given by the simple roots $\vec{\alpha}_n$ ($n=1,\ldots, N-1$). They correspond to the 't~Hooft-Polyakov monopoles~\cite{tHooft:1974kcl, Polyakov:1974ek}.
The last one~\cite{Lee:1997vp, Lee:1998bb, Kraan:1998kp, Kraan:1998pm, Kraan:1998sn} is called the Kaluza–Klein (KK) monopole, which carries the fractional topological charge $1/N$ and the magnetic charge given by the Affine simple root, $\vec{\alpha}_N=-\vec{\alpha}_1-\cdots - \vec{\alpha}_{N-1}$. 
With dual photons, these vertices take the form of
\begin{equation}
    \zetam\, \rme^{\im \vec{\alpha}_n\cdot \vec{\sigma}+\im \theta/N}
    \quad (n=1,\ldots,N),
    \label{eq:3dMonopoleVertex}
\end{equation}
where $\theta$ is the vacuum angle and $\zetam=O(\rme^{-S_I/N})$ is the monopole fugacity mainly controlled by the instanton action $S_I=\frac{8\pi^2}{g^2}$. 
The $3$d effective Lagrangian is obtained by their dilute gas approximation as in the case of the Polyakov model~\cite{Polyakov:1975rs}, and we find~\cite{Unsal:2007vu, Unsal:2007jx, Unsal:2008ch, Shifman:2008ja, Poppitz:2012nz, Poppitz:2012sw}
\begin{equation}
    \mathcal{L}_{3\mathrm{d}}= \frac{g^2}{16\pi^2 L_4}|\diff \vec{\sigma}|^2 - \sum_{n=1}^{N} 2\zetam \cos \left(\vec{\alpha}_n\cdot \vec{\sigma}+\frac{\theta}{N}\right). 
    \label{eq:3dEffectiveLagrangian}
\end{equation}
The monopole gas produces the nonperturbative mass gap for dual photons, and this explains the area law for the spatial Wilson loops.

\section{Confinement on \texorpdfstring{$\mathbb{R}^2\times T^2$}{R2xT2} and vortex}

We consider compactifying another direction and put the $4$d YM theory on $\mathbb{R}^2\times T^2\ni (\bm{x}, (x_3,x_4))$ with $x_3\sim x_3+L_3$ and $x_4\sim x_4+L_4$. 
Along the $T^2$ direction, we take the 't~Hooft twisted boundary condition~\cite{tHooft:1979rtg}, so that the classical vacuum is given by the flat gauge field with non-commuting holonomies, $P_3 P_4 = P_4 P_3 \rme^{2\pi \im/N}$. 
Unlike the $3$d case, the classical vacuum is center symmetric without introducing the double-trace deformation thanks to the boundary condition, and it is given by the $SU(N)$ shift and clock matrices~\cite{Tanizaki:2022ngt}: Up to gauge transformations,
\begin{equation}
    P_3 \propto S , \quad P_4 \propto C, 
\end{equation}
where $(S)_{ij}= \delta_{i+1,j}$. The low-energy gauge group of the $2$d effective theory becomes $SU(N)\xrightarrow{\mathrm{Higgs}}\mathbb{Z}_N$, and the perturbative spectrum is already gapped.

In this setup, the $4$d instanton splits into $N$ identical center vortices~\cite{Gonzalez-Arroyo:1998hjb, Montero:1999by, Montero:2000pb, Gonzalez-Arroyo:2023kqv}. 
The center vortex carries the fractional topological charge $1/N$ and has the nontrivial commutation relation with the $2$d Wilson loops. 
Due to the perturbative mass gap, the center vortex has a fixed size, which justifies the dilute gas approximation. The partition function can be computed as~\cite{Tanizaki:2022ngt}
\begin{align}
    Z_{2\mathrm{d}}&= \sum_{n,\overline{n}}\frac{\delta_{n-\overline{n}\in N \mathbb{Z}}}{n! \overline{n}!} (V \zetav \rme^{\im \theta/N})^n(V \zetav \rme^{-\im \theta/N})^{\overline{n}} \notag\\
    &=\sum_{k=1}^{N}\exp\left[ 2V \zetav \cos\left(\frac{\theta+2\pi k}{N}\right)\right],  \label{eq:center-vortex-gas}
\end{align}
where $\zetav=O(\rme^{-S_I/N})$ is the center-vortex fugacity, $V$ is the volume of $\mathbb{R}^2$, and $n$ and $\overline{n}$ represent numbers of vortex and anti-vortex, respectively. $\delta_{n-\overline{n}\in N\mathbb{Z}}=\sum_{k=1}^{N}\rme^{\frac{2\pi \im}{N} k (n-\overline{n})}$ is required to represent the integer-quantization of the total topological charge, $Q_{\mathrm{top}}=\frac{n-\overline{n}}{N}$. 
This produces the multi-branch structure of the $\theta$ vacua, and the ground-state energy with the label $k$ behaves as
\begin{equation}
    E_k(\theta)= -2\zetav \cos\left(\frac{\theta+2\pi k}{N}\right). 
    \label{eq:multibranch_vortex}
\end{equation}
One can also derive the area law for the Wilson loops inside $\mathbb{R}^2$~\cite{Tanizaki:2022ngt}.

\section{Derivation of the center-vortex gas from the monopole gas}

As we have seen, there are two distinct weakly-coupled descriptions for the confinement of $4$d YM theory in compactified geometries: One of them is described by monopoles on $\mathbb{R}^3\times S^1$ and the other is by center vortex on $\mathbb{R}^2\times T^2$ with the 't~Hooft flux. 
Both are constituents of the instanton, but they have different magnetic features. 
For example, there are $N$ different types of monopoles on $\mathbb{R}^3\times S^1$ and all of them contribute equally to the nonperturbative potential, while we only have a single type of vortex on $\mathbb{R}^2 \times T^2$. 
We are going to unify these two semiclassical confinement mechanisms.

We consider the case $L_3\gg L_4$ 
and add the double-trace deformation for $P_4$, 
so that the $3$d effective theory~\eqref{eq:3dEffectiveLagrangian} is applicable. 
As the 't~Hooft flux is important to achieve the $2$d semiclassical description, we first clarify how it affects dual photons. 

$4$d YM theory enjoys the $\mathbb{Z}_N$ $1$-form symmetry~\cite{Gaiotto:2014kfa}, denoted as $\mathbb{Z}_N^{[1]}$. 
The 't~Hooft flux corresponds to the symmetry-twisted boundary condition of $\mathbb{Z}_N^{[1]}$, and it plays the pivotal role to maintain the $4$d 't~Hooft anomalies in the $2$d effective theories~\cite{Tanizaki:2022ngt, Tanizaki:2017qhf, Tanizaki:2017mtm, Yamazaki:2017dra, Dunne:2018hog}. 
When putting the theory on $\mathbb{R}^3\times S^1$, the $4$d $1$-form symmetry becomes 
\begin{equation}
    (\mathbb{Z}_N^{[1]})_{4\mathrm{d}}\xrightarrow{\mathbb{R}^3\times S^1} (\mathbb{Z}_N^{[1]})_{3\mathrm{d}}\times (\mathbb{Z}_N^{[0]})_{3\mathrm{d}},  
    \label{eq:3dSymmetry}
\end{equation}
where the $3$d $1$-form symmetry acts on the spatial loops while the $0$-form symmetry acts on $P_4$, $P_4\mapsto \omega P_4$.
To realize $\mathbb{R}^2\times T^2$, we regard it as the compactification, $\mathbb{R}^3\Rightarrow \mathbb{R}^2\times S^1 \ni (\bm{x},x_3)$, and then the $(\mathbb{Z}_N^{[1]})_{4\mathrm{d}}$-twisted boundary condition on $\mathbb{R}^2\times T^2$ should correspond to the $(\mathbb{Z}_N^{[0]})_{3\mathrm{d}}$-twisted boundary condition on $\mathbb{R}^2\times S^1$ for \eqref{eq:3dEffectiveLagrangian}. 

The $(\mathbb{Z}_N^{[0]})_{3\mathrm{d}}$ transformation in \eqref{eq:3dEffectiveLagrangian} is realized as~\cite{Anber:2015wha}  
\begin{equation}
    \vec{\sigma} \mapsto S^{-1} \vec{\sigma}. 
    \label{eq:3d0formSymmetry}
\end{equation}
To see this, we note that the symmetry transformation $P_4\mapsto \omega P_4$ is equivalent to $P_4\to S P_4 S^{-1}$ at the classical vacuum~\eqref{eq:VEVofP4}. 
Thus, we can combine the $0$-form symmetry and the gauge transformation by the shift matrix $S$ to make the expectation value of $P_4$ unchanged, but the dual photon is now transformed by $S$. 
Therefore, the $(\mathbb{Z}_N^{[0]})_{3\mathrm{d}}$ symmetry permutes the monopole vertices, 
\begin{equation}
    \rme^{\im \vec{\alpha}_n\cdot \vec{\sigma}+\im \theta/N} 
    \mapsto 
    \rme^{\im \vec{\alpha}_n\cdot S^{-1}\vec{\sigma}+\im \theta/N} 
    = \rme^{\im \vec{\alpha}_{n+1}\cdot \vec{\sigma}+\im \theta/N}. 
\end{equation}
The effective Lagrangian~\eqref{eq:3dEffectiveLagrangian} enjoys the $\mathbb{Z}_N$ symmetry because the BPS and KK monopoles equally contribute.

Now, it becomes evident that the 't~Hooft twisted boundary condition in view of \eqref{eq:3dEffectiveLagrangian} is given by 
\begin{equation}
    \vec{\sigma}(\bm{x},x_3+L_3)= S^{-1} \vec{\sigma}(\bm{x},x_3). 
    \label{eq:TwistedBC}
\end{equation}
This boundary condition eliminates the zero-modes, and the dual photons get the perturbative gap of $2\pi/NL_3$, reflecting the fact that the perturbative spectrum on $\mathbb{R}^2\times T^2$ with the 't~Hooft flux is gapped. 
The constant modes of $\vec{\sigma}$ are restricted to the $N$ distinct vacua, 
\begin{equation}
    \Vec{\sigma}^*_k = \frac{2 \pi k}{N} \vec{\rho}, 
    \label{eq:ConstantMode_DualPhoton}
\end{equation}
where $k=0,1,\ldots,N-1$ and $\vec{\rho}=\Vec{\mu}_1 + \cdots + \Vec{\mu}_{N-1}$ is the Weyl vector. 
Thus, the low-energy gauge group is further reduced as $U(1)^{N-1}\xrightarrow{\mathrm{Higgs}} \mathbb{Z}_N$. 
Let us point out that $\vec{\sigma}=\vec{\sigma}^*_k$ are the saddle points of the monopole potential~\cite{Unsal:2007jx}: The nonperturbative vacua of~\eqref{eq:3dEffectiveLagrangian} at large $L_3$ and the classical vacua at small $L_3$ are smoothly connected.

Let us retrieve the center-vortex gas partition function from the monopole gas description. 
Note that the Weyl vector satisfies $\Vec{\alpha}_n \cdot \Vec{\rho} = 1~(\operatorname{mod} N)$
for $n = 1, \cdots, N$ including the Affine root.
For the classical vacua $\Vec{\sigma} = \Vec{\sigma}^*_k$, the BPS and KK monopoles become an identical vertex,
\begin{equation}
    \rme^{\im \vec{\alpha}_n\cdot \vec{\sigma}+\im \theta/N}  
    \Rightarrow \rme^{\im (2\pi k+\theta)/N} ,
\end{equation}
and we identify it as the center-vortex vertex of the $2$d effective theory. 
When $L_3\,(\,\gg L_4)$ is much smaller than the strong scale, we can reduce the path integral of $\vec{\sigma}$ to the summation over the classical vacua $\Vec{\sigma} = \Vec{\sigma}^*_k$, so the $3$d partition function becomes
\begin{align}
    Z_{3\mathrm{d}} &= \int \mathcal{D}\vec{\sigma} ~\rme^{- \int \mathcal{L}_{3d}} \approx \sum_{\vec{\sigma} = \Vec{\sigma}^*_k } ~\rme^{- \int \mathcal{L}_{3d}} \notag \\
    &= \sum_{k \in \mathbb{Z}_N} \exp\left[ 2 N V L_3 \zetam \cos\left(\frac{\theta+2\pi k}{N}\right)\right].
    \label{eq:multibranch_monopole}
\end{align}
This reproduces the center-vortex gas formula~\eqref{eq:center-vortex-gas} by setting $\zetav=NL_3\zetam$.
This observation connects the two semiclassical descriptions.

\section{Monopoles become the center vortex}

\begin{figure}[t]
\includegraphics[width = 0.9 \linewidth]{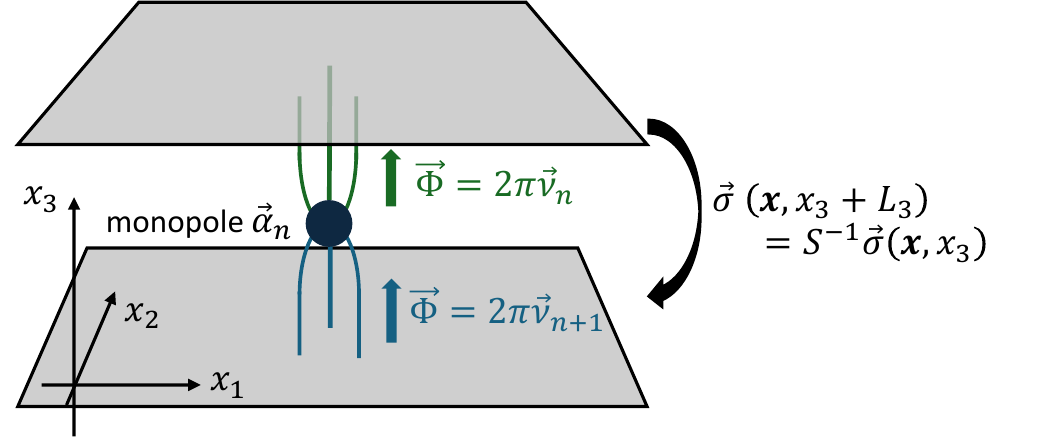}
\caption{Schematic illustration of the magnetic field emitted from the monopole in the shift-twisted $\mathbb{R}^2 \times S^1$ setup. Due to the perturbative mass gap, the magnetic flux is squeezed into the region $|\bm{x}|\lesssim \frac{N L_3}{2\pi}$, and the $\vec{\alpha}_n$-monopole carries the $2\pi\vec{\nu}_n$ outgoing flux and the $2\pi\vec{\nu}_{n+1}$ incoming flux along the $S^1$ direction. 
This becomes the center vortex of the $2$d description.}
\label{fig:monopole-vortex}
\end{figure}

In what follows, we examine the monopole vertex in $\mathbb{R}^2 \times S^1$ microscopically and illustrate how different fundamental monopoles become the identical center vortex.
It is convenient to introduce the weight vectors for the $SU(N)$ defining representation, 
\begin{equation}
    \vec{\nu}_1=\vec{\mu}_1,\, \vec{\nu}_2=\vec{\mu}_1-\vec{\alpha}_1,\,\ldots, \, \vec{\nu}_N=\vec{\mu}_1-\vec{\alpha}_1-\cdots - \vec{\alpha}_{N-1}, \label{eq:weight_vector_def}
\end{equation}
and the Affine roots can be expressed as $\vec{\alpha}_n=\vec{\nu}_{n}-\vec{\nu}_{n+1}$ for $n=1,\ldots, N$. 
Then, we may regard the $\vec{\alpha}_n$-monopole carries the incoming $2\pi\vec{\nu}_{n+1}$ magnetic flux and the outgoing $2\pi\vec{\nu}_{n}$ magnetic flux.
As the shift-twisted boundary condition converts the $2\pi \vec{\nu}_{n}$ magnetic flux to the $2\pi \vec{\nu}_{n+1}$ magnetic flux, 
we can consider a vortex-like configuration of the monopole, where the magnetic flux forms a tube along the $S^1$ direction. 
This is schematically shown in Fig.~\ref{fig:monopole-vortex} (see also~\cite{Unsal:2020yeh}). 
We shall show that this solves the classical equation of motion in the dual-photon theory.

For the $\vec{\alpha}_n$-monopole at $(\bm{x},x_3)=(\bm{0}, x_*)$, the classical equation of motion for $\vec{\sigma}$ becomes 
\begin{equation}
    \nabla^2  \left(\frac{g^2}{4\pi \im L_4}\vec{\sigma}\right)= 2\pi \vec{\alpha}_n\, \delta(\bm{x})\delta(x_3-x_*).  
\end{equation}
Extending $x_3\in \mathbb{R}/L_3 \mathbb{Z}$ to $x_3\in \mathbb{R}$, the $(\mathbb{Z}_N^{[0]})_{3\mathrm{d}}$-twisted boundary condition~(\ref{eq:TwistedBC}) puts extra monopoles with permuting their magnetic charges as $x_3$ is shifted by $L_3$: 
\begin{equation}
    \nabla^2 \left(\frac{g^2}{4\pi \im L_4}\vec{\sigma}\right) = 2\pi \sum_{k\in \mathbb{Z}} \vec{\alpha}_{n-k}\, \delta(\bm{x})\delta(x_3-x_*-kL_3), 
    \label{eq:MagneticCoulombEq}
\end{equation}
where the index for $\vec{\alpha}_{n-k}$ is understood in mod $N$. 
The solution of \eqref{eq:MagneticCoulombEq} with $\vec{\sigma}(\bm{x},x_3)\to 0$ at $|\bm{x}|\to \infty$ is
\begin{align}
    \frac{g^2}{4\pi \im L_4}\vec{\sigma}&=-\frac{1}{2}\sum_{\ell=0}^{N-1}\vec{\nu}_{n-\ell} \sum_{k\in \mathbb{Z}}\left(\frac{1}{\sqrt{|\bm{x}|^2+(x_{3,\ell}- NkL_3)^2}}\right.\notag\\
    &\quad\quad\quad \left.-\frac{1}{\sqrt{|\bm{x}|^2+(x_{3,\ell}-L_3- NkL_3)^2}}\right), 
    \label{eq:SolutionMagneticPotential}
\end{align}
where $x_{3,\ell}=x_3-x_*-\ell L_3$. 
We can see that $\vec{\sigma}=O(\rme^{-\frac{2\pi}{N L_3}|\bm{x}|})$ as $|\bm{x}|\to \infty$, and thus the magnetic flux emitted from the monopole is squeezed into the bounded region $|\bm{x}|\lesssim \frac{N L_3}{2\pi}$, which forms the vortex of $2$d theory. 

We can obtain the vortex magnetic flux by evaluating the magnetic flux $\vec{\Phi}$ for the constant $x_3$ plane. Following the definition~\eqref{eq:DualPhotonDef}, we have\footnote{There is a subtlety on interchanging the order of the sum and integral. However, the additional term arising from this subtlety eventually vanishes due to $\sum_{\ell=0}^{N-1}\vec{\nu}_{\ell} = 0$, so we just drop it here.}
\begin{align}
    \vec{\Phi}(x_3)&=\int_{\mathbb{R}^2} \diff^2\bm{x}\,\frac{g^2}{4\pi \im L_4}\partial_{x_3}\vec{\sigma}(\bm{x},x_3) \notag\\
    &=\pi\sum_{\ell=0}^{N-1}\vec{\nu}_{n-\ell} \sum_{k\in \mathbb{Z}}\Big[\mathrm{sign}(x_{3,\ell}-NkL_3) \notag\\
    &\quad\quad\quad\quad\quad\quad -\mathrm{sign}(x_{3,\ell}-L_3-Nk L_3)\Big]. 
\end{align}
When restricting our attention to the original domain $0\le x_3<L$, we find that 
\begin{equation}
    \vec{\Phi}(x_3)=\left\{
    \begin{array}{cc}
         2\pi \vec{\nu}_{n}& (x_*<x_3<L_3), \\
         2\pi \vec{\nu}_{n+1}& (0\le x_3<x_*). 
    \end{array}
    \right.
\end{equation}
This is exactly the vortex configuration shown in Fig.~\ref{fig:monopole-vortex}. 
Now, $x_3$ becomes the internal moduli of the vortex in the $2$d perspective, and there is no longer a distinction between BPS and KK monopoles as they are permuted by $x_3\mapsto x_3+L_3$. 
This explains why we have the unique type of the center vortex on $\mathbb{R}^2\times T^2$ with the 't~Hooft flux although there are $N$ types of fundamental monopoles.

We conclude this section by showing that this vortex configuration is actually the center vortex: The phase of the Wilson loop is rotated by a center element depending on the presence of this vortex.
With the 3d $U(1)^{N-1}$ gauge field $\vec{a}$, the $SU(N)$ fundamental Wilson loop reads
\begin{equation}
    W(C) = \frac{1}{N}\sum_{\ell' =1}^N \exp\left({\im \vec{\nu}_{\ell'} \cdot \int_C \vec{a}}\right).
\end{equation}
We take $C$ to be the loop inside $\mathbb{R}^2\subset \mathbb{R}^2\times S^1$, 
and we describe it as the boundary of a surface, $C=\partial \Sigma$. 
Then, 
\begin{align}
    \int_C \vec{a}&=\int_{\Sigma}\diff \vec{a} = \int_{\Sigma} \frac{g^2}{4\pi \im L_4}\star \diff \vec{\sigma} \notag\\
    &=\left\{
    \begin{array}{cc}
       2\pi \vec{\nu}_n  & (\text{vortex lies inside of $C$}), \\
       0  & (\text{vortex lies outside of $C$}).
    \end{array}
    \right.
\end{align}
The label of $\vec{\nu}_n$ depends on the $x_3$ location of the surface $\Sigma$, but it does not affect the following result. 
Note that $\vec{\nu}_n\cdot \vec{\nu}_m=\delta_{nm}-\frac{1}{N}$, and thus the phase of the Wilson loop is determined as 
\begin{align}
    W(C)&=\left\{
    \begin{array}{cc}
       \rme^{-2\pi \im/N}  & (\text{vortex lies inside of $C$}), \\
       1  & (\text{vortex lies outside of $C$}).
    \end{array}
    \right.
\end{align}
This is exactly the characterization of the center vortex.

\section{Conclusion and outlook}

The liberation of magnetic objects is the key ingredient to explain quark confinement, and magnetic monopoles and center vortices are two promising candidates for such excitations. 
This work has established a concrete connection between monopoles and center vortices within the analytically controllable semiclassical regime~(Fig.~\ref{fig:monopole-vortex}), and we unified the $3$d monopole theory on $\mathbb{R}^3\times S^1$ and the $2$d center-vortex theory on $\mathbb{R}^2\times T^2$ with the 't~Hooft flux.
Our finding provides a proof of concept for the renowned scenario~\cite{DelDebbio:1997ke, Ambjorn:1999ym, deForcrand:2000pg} that the monopole serves as the kink or endpoint of the center-vortex network.

In the 4d center-vortex scenario of confinement, the role of monopole makes the center-vortex worldsheets non-orientable, which has the significant effect on the topological charge~\cite{Engelhardt:1999xw, Reinhardt:2001kf, Cornwall:1999xw}. 
In the center-vortex picture, the topological charge density of each configuration localizes at the intersection point of center vortex worldsheets, and each intersection point gives the fractional topological charge quantized in $1/N$. 
We note that the confinement vacuum must have the multi-branch structure due to the mixed 't~Hooft anomaly between $\mathbb{Z}_N^{[1]}$ and the $\theta$ periodicity~\cite{Gaiotto:2017yup}. The fractional topological charge at the intersection of center-vortex worldsheets can naturally explain its origin, and one may regard Eq.~\eqref{eq:multibranch_vortex} as its $2$d semiclassical realization. 
Without monopoles, however, the total topological charge becomes $0$ as the net intersection number vanishes due to the cancellation between intersections with opposite orientation. 
Therefore, the $\theta$ dependence requires non-oriented center-vortex worldsheets by locating monopoles at the junction.
In addition, recent studies~\cite{Oxman:2018dzp, Junior:2019fty, Junior:2022bol} suggest that the nonoriented vortex worldsheets explain the asymptotic behaviors of string tensions for various representations, keeping the Abelian-like flux-tube profile.

The ideas presented in this Letter open up diverse directions for future works, which would deepen our understanding of the confinement mechanism. 

Firstly, we anticipate the fruitful interplay between the monopole and center-vortex semiclassical frameworks. 
For instance, the 3d monopole theory for adjoint QCD is more developed than its 2d center-vortex description. 
Conversely, for QCD with fundamental quarks, the monopole semiclassics cannot treat the non-Abelian nature of chiral symmetry~\cite{Poppitz:2013zqa, Cherman:2016hcd}, whereas the center-vortex one respects the full chiral symmetry and also uncovers properties of the $\eta'$~\cite{Tanizaki:2022ngt, Hayashi:2024qkm}. Both frameworks can offer valuable insights to each other.

Secondly, the analytic solution describing the center vortex is not yet known, and this is a crucial missing piece for the $2$d center-vortex theory.
In this Letter, the explicit solution of the magnetic potential (\ref{eq:SolutionMagneticPotential}) has been obtained by aligning the BPS and KK monopoles and 
this idea could enable us to construct an analytic solution of the center vortex in the $ \mathbb{R}^2 \times T^2$ setup.

Lastly, we aspire this work will advance a unified comprehension of the center-vortex and dual superconductor paradigms, and it would be valuable to re-examine the monopole-versus-center-vortex discussion.

\section*{Acknowledgements}
The authors thank Tatsuhiro Misumi, Erich Poppitz, and Mithat \"Unsal for useful discussions.
This work was partially supported by Japan Society for the Promotion of Science (JSPS) Research Fellowship for Young Scientists Grant
No. 23KJ1161 (Y.H.), by JSPS KAKENHI Grant No. 23K22489 (Y.T.), and by Center for Gravitational Physics and Quantum Information (CGPQI) at Yukawa Institute for Theoretical
Physics.

\bibliographystyle{utphys}
\bibliography{./refs.bib,./QFT.bib}
\end{document}